# Experimental demonstration of complete 180° reversal of magnetization in isolated Co-nanomagnets on a PMN-PT substrate with voltage generated strain

Ayan Kumar Biswas[1], Hasnain Ahmad[1], Jayasimha Atulasimha[2] and Supriyo Bandyopadhyay[1*]

[1]Department of Electrical and Computer Engineering,

[2]Department of Mechanical and Nuclear Engineering

Virginia Commonwealth University,

Richmond, VA 23284, USA

**Rotating the magnetization of a shape anisotropic magnetostrictive nanomagnet with voltage-generated stress/strain dissipates much less energy than most other magnetization rotation schemes, but its application to writing bits in non-volatile magnetic memory has been hindered by the fundamental inability of stress/strain to rotate magnetization by full 180°. Normally, stress/strain can rotate the magnetization of a shape anisotropic elliptical nanomagnet by only up to 90°, resulting in incomplete magnetization reversal. Recently, we predicted that applying uniaxial stress sequentially *along two different axes* that are not collinear with the major or minor axis of the elliptical nanomagnet will rotate the magnetization by full 180° [1]. Here, we demonstrate this complete 180° rotation in elliptical Co-nanomagnets (fabricated on a piezoelectric substrate) at room temperature. The two stresses are generated by sequentially applying voltages to two pairs of shorted electrodes placed on the substrate such that the line joining the centers of the electrodes in one pair intersects the major axis of a nanomagnet at ~+30° and the line joining the centers of the electrodes in the other pair intersects at ~ -30°. A finite element analysis has been performed to determine the stress distribution underneath the nanomagnets when one or both pairs of electrodes are activated, and this has been approximately incorporated into a micromagnetic simulation of magnetization dynamics to confirm that the generated stress can produce the observed magnetization rotations. This result portends an extremely energy-efficient non-volatile "straintronic" memory technology predicated on writing bits in nanomagnets with electrically generated stress.**

Nanomagnets are the bedrock of non-volatile memory. A magnetic random access memory (MRAM) cell is implemented with a magneto-tunneling junction (MTJ) consisting of two nanomagnetic layers, one hard and one soft, separated by a spacer (tunneling) layer. The soft layer is often shaped like an elliptical disc which, if sufficiently thick, has two in-plane stable magnetization directions pointing in opposite directions along the major axis of the ellipse. They encode and store the binary bits 0 and 1. The stored bit is "read" by measuring the resistance of the MTJ which has two discrete values depending on the two magnetization orientations of the soft layer, i.e., for the two bits 0 and 1. "Writing" of bits is accomplished by switching the magnetization of the soft layer between the two anti-parallel directions of stable magnetization ($180^0$ rotation of the magnetization) with an external agent.

---

[*] Corresponding author. E-mail: sbandy@vcu.edu

There are many strategies to rotate the magnetization of the soft layer. Popular approaches include passing a spin current through the soft layer to generate a spin transfer torque[2-7] or spin orbit torque[8-11] or domain wall motion[12-13]. Other approaches involve using voltage controlled magnetic anisotropy[14], magnetoelectric effects[15-17], magnetoionic effects[18] and magnetoelastic effects[19-25]. Unfortunately, generation of a spin current requires passing a charge current through a resistor that dissipates excessive energy, making the spin-current based schemes relatively energy-inefficient[26, 27]. The voltage based methods also dissipate energy in charging a capacitor, but turn out to be more energy-efficient. One magnetoelastic scheme, the so-called "straintronic" switching, involves rotating the magnetization of a magnetostrictive soft layer with mechanical strain generated by applying a voltage across an underlying piezoelectric layer with a suitable arrangement of electrodes[28-30]. The voltage generates strain in the piezoelectric, which is partially or completely transferred to the elliptical magnetostrictive soft layer, and rotates magnetization by the Villari effect[19-25]. It has been predicted theoretically that large angle (~$90^o$) rotation in ~100 nm feature sized nanomagnets made of highly magnetostrictive materials (Terfenol-D, FeGa) will dissipate only ~1 aJ of energy to occur in ~1 ns[31-33]. Recent experiments[34-26] have confirmed that the energy dissipated to switch a nanomagnet in this fashion will be on the order of 1 aJ in properly scaled structures.

Despite the excellent energy-efficiency, straintronic switching is not used for writing bits in MRAMs. What has prevented its use is that strain can normally rotate the magnetization of an elliptical nanomagnet by only up to $90^o$ from either stable orientation along the major axis, placing the final magnetization state along the unstable minor (magnetically hard) axis. Upon strain withdrawal, the magnetization will return to one of the two stable orientations along the major axis, but not necessarily the desired orientation. It has *equal probability* of reaching either orientation. This allows for writing of bits with only 50% success probability, which, of course, is unacceptable. Recently, we proposed a practical approach to increasing the success probability to well over 99.9999% at room temperature in the presence of thermal noise[1]. The trick is to apply uniaxial stresses sequentially in two different directions, non-collinear with the major or minor axis, by activating in succession two pairs of electrodes delineated on the piezoelectric's surface. The two members of each pair are electrically shorted with each other and either pair is activated by applying a voltage between it and the grounded bottom of the substrate. The electrodes are arranged such that the line joining the centers of the members of one pair subtends an angle of +$30^o$, and the line joining the members of the other pair subtends an angle of -$30^o$, with one direction along the ellipse's major axis. Sequential stressing rotates the magnetization vector in two steps. In the first step, the magnetization rotates from the initial stable orientation along the major axis by an acute angle. In the second step, the magnetization rotates by an additional angle, bringing it closer to the other stable direction, and finally when the stresses are withdrawn, the magnetization settles into the other stable direction, completing a $180^o$ rotation[1].

There are other proposed methods of implementing $180^o$ rotation with strain, but they either require very precise timing of the stress cycle which is nearly impossible in the presence of thermal noise at room temperature[37, 38], or special material properties[39]. In contrast, the two step method does not call for extreme precision, is practical and error-resilient, and works with any magnetostrictive material, whether crystalline, poly-crystalline or amorphous.

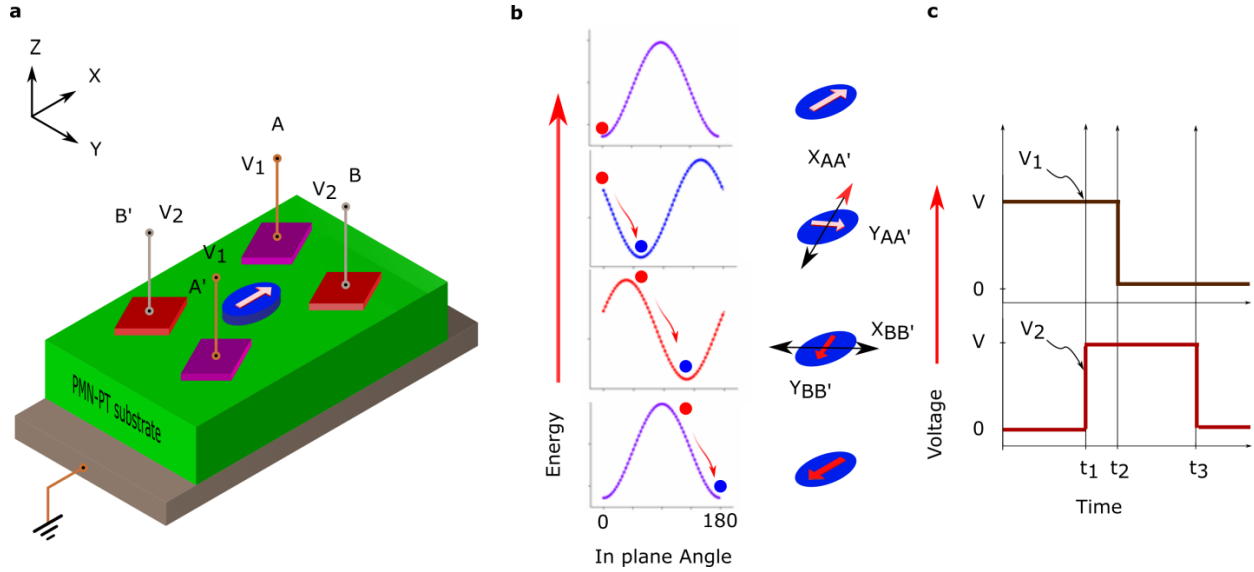

Fig. 1: **Strain-induced complete magnetization reversal scheme**. (a) An elliptical Co nanomagnet is fabricated at the intersection of the lines joining the centers of two pairs of electrodes $AA'$ and $BB'$ delineated on a poled piezoelectric $Pb(Mg_{1/3}Nb_{2/3})O_3$-$PbTiO_3$ (PMN-PT) substrate whose bottom is grounded. The substrate is poled with an electric field in the $z$-direction. The lines $AA'$ and $BB'$ subtend angles of $+30°$ and $-30°$, respectively, with the major axis of the elliptical nanomagnet. (b) The potential energy of the nanomagnet as a function of the angle that its magnetization subtends with the major axis is shown in the left panel for four stressing scenarios - neither electrode pair is activated with a voltage, only $AA'$ is activated, only $BB'$ is activated, and both are deactivated. The red dot denotes the initial orientation of the magnetization and the blue dot the final orientation in any stressing scenario. The final orientation always conforms to an energy minimum. The right panel shows the magnetization orientations corresponding to the minima in the corresponding potential energy profiles in the left panel, i.e. they are the stable orientations in the four different stressing scenarios. (c) The timing diagram of the voltage pulses at the two electrode pairs. Figures are not drawn to scale.

Fig. 1 shows a cartoon of the 180° rotation scheme. A single elliptical Co nanomagnet is delineated on the surface of a poled piezoelectric substrate. Two pairs of electrodes are placed on the substrate's surface such that the line joining the centers of the electrodes in the first pair $AA'$ makes an angle of $+30°$ with one direction along the major axis of the nanomagnet and the line joining the centers of the electrodes in the second pair $BB'$ makes an angle of $-30°$ (or equivalently $+330°$) with the same direction. The bottom of the substrate is grounded. The direction of substrate poling is such that when a positive voltage $V_1$ (= $V$) is applied between the electrically shorted first pair $AA'$ and ground, the nanomagnets are elongated along the line $X_{AA'}$ joining the electrode pair and are contracted along the line $Y_{AA'}$ in the perpendicular direction (see the right vertical panel of Fig. 1b)[28]. A tensile strain $\varepsilon_{xx}$ is generated in the nanomagnet along $X_{AA'}$ and a compressive strain $\varepsilon_{yy}$ is generated along the line $Y_{AA'}$. An effective tensile strain $\varepsilon_{xx}$-$\varepsilon_{yy}$ generated along $X_{AA'}$. Since Co has a negative magnetostriction coefficient, the nanomagnet's magnetization rotates through an acute angle from the initial orientation shown in Fig. 1a and aligns along the direction perpendicular to the strain axis, $Y_{AA'}$ (see the right panel of Fig. 1b). The potential energy profile in the plane of the magnet, drawn as a function of the angle that the in-plane component of the magnetization subtends with the initial direction along the major axis is shown in the left panel of Fig. 1b for the four different stressing scenarios: no electrode is activated with a voltage, only *AA'* is activated, only *BB'* is activated and both electrodes are deactivated. Note that activating *AA'* places the magnetization at an acute angle from the initial orientation. Next, a positive voltage $V_2$ (= $V_1$ = $V$) is applied between the electrically shorted second pair *BB'* and ground at time $t_1$ while turning off the voltage ($V_1$) at pair *AA'* after a time $t_2$ ($t_2 \geq t_1$ ). This will again generate an effective tensile strain ($\varepsilon_{xx}$-$\varepsilon_{yy}$) along the line $X_{BB'}$ and the magnetization rotates further and aligns in the direction $Y_{BB'}$ perpendicular to that of the new strain axis along *BB'*. The new energy profile in the plane of the magnet (third row of Fig. 1b) shows that the magnetization has rotated through an obtuse angle from the initial orientation shown in the first row. Finally, the voltage ($V_2$) is turned off at time $t_3$ ($t_3 > t_1, t_2$ ), whereupon the bistable energy profile is restored and the magnetization settles down in the direction that is *opposite* to the initial direction since it cannot transcend the energy barrier that separates the last orientation from the initial one[1]. This completes a 180° rotation. The same rotation would occur if we reversed the sequence of stress application from (*AA'*, *BB'*) to (*BB'*, *AA'*), except that in one case, the rotation will take place clockwise and in the other case counter-clockwise. Note that if we activated only one pair of electrodes – either *AA'* or *BB'* – and not the other, then the magnetization will have rotated through an acute angle from the initial orientation and finally returned to the initial orientation after electrode deactivation, resulting in 0° rotation.

Note also that there are no restrictions on the time durations $t_1$, $t_2$ and $t_3$, except that they must exceed the times it takes for the magnetization to rotate between the intermediate states. Since the latter are uncertain in the presence of thermal noise, we merely need to ensure that $t_1$, $t_2$ and $t_3$ are larger than the statistically largest possible rotational durations. Thus, there is no critical precision demand on $t_1$, $t_2$ or $t_3$, which makes the scheme practical.

To demonstrate this scheme experimentally, we have fabricated the structure of Fig. 1a, except that we made large area contact pads and placed several nanomagnets between them instead of a single nanomagnet. This was done to make the lithography tractable without compromising the demonstration. The nanomagnets are arranged with their major axes mutually parallel (within lithographic tolerance), such that the lines *AA'* and *BB'* joining the electrode pairs subtend angles of $\pm 30^0$ with one direction of the major axis. All nanomagnets are initially magnetized along one direction of the major axes with a global magnetic field of 0.2 Tesla. The magnetization states of the nanomagnets before and after application of voltage generated stresses are determined by magnetic force microscopy (MFM) using a low moment tip.

Fig. 2a shows AFM images of four nanomagnets which are ~700 nm apart from each other (center-to-center distance between nearest neighbors). The large inter-magnet distance ensures that dipole interactions between nanomagnets are negligible and each nanomagnet can be viewed as "isolated" from its neighbors. The major and minor axes of the nanomagnets are measured from the AFM image to be 198 nm and 183 nm, respectively. The nominal nanomagnet thickness is 8 nm, as determined during deposition, but we will assume it to be 7 nm to allow for 1 nm of surface oxidation.

With these dimensions, the in-plane shape anisotropy energy barrier in a nanomagnet is calculated to be 4.191 eV. It is not so large that stress cannot overcome it in the manner of Fig. 1b and make the magnetization switch, and yet not so small that the magnetized MFM tip could alter the magnetization during scanning. Fig. 2b shows the MFM images of these nanomagnets after magnetization by the global magnetic field and before application of any stress. It is evident from Fig. 2b that the magnetic field aligns all but one nanomagnet's magnetization more or less along the field. They are not exactly along the field since the magnetization will seek out the closest easy axis, which may not be exactly along the major axis because the magnet shapes are not perfectly elliptical. However, the outlier nanomagnet in the top left corner is magnetized almost opposite to the magnetic field. Its failure to align its magnetization along or close to the magnetic field can be ascribed to a variety of effects, but most likely is caused by pinning of the magnetization in the direction almost opposite to the applied field. Note that all nanomagnets exhibit near single domain characteristics with distinct N-S poles.

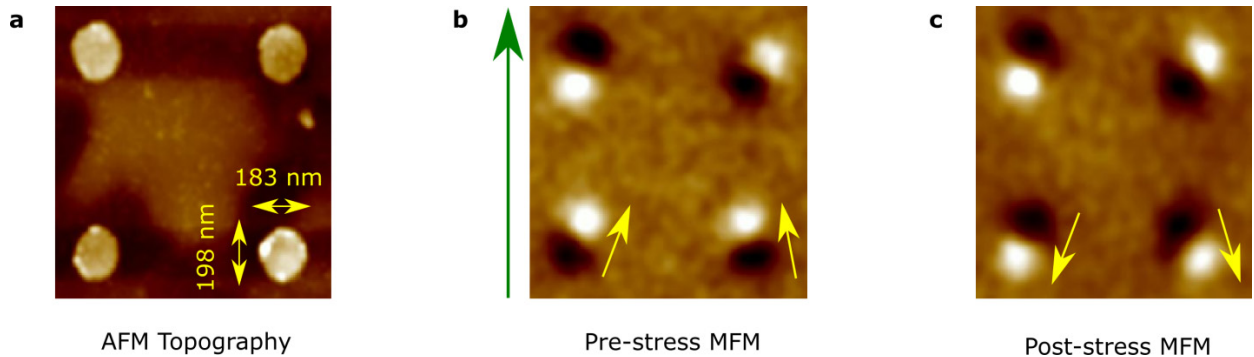

Fig. 2: **Atomic force micrographs (AFM) and magnetic force micrographs (MFM) of four nominally identical elliptical Co nanomagnets delineated on a PMN-PT substrate showing how their magnetizations evolve with stress**. (a) AFM image showing the topography of the nominally elliptical nanomagnets. (b) The nanomagnets are magnetized with a high magnetic field (~0.2 Tesla) in the direction of the green arrow (approximately parallel to the major axes of the ellipses) prior to applying stress. After field withdrawal, each nanomagnet's magnetization rotates to the nearest easy axis (energy minimum) which may not be exactly along the major axis of the ellipse because the magnets' shapes are not perfectly elliptical. The top left nanomagnet's magnetization was unaffected by the field; it is pinned in a stable single domain state that is magnetized nearly anti-parallel to the applied field. The pinning can be caused by lithographic imperfections, zagged edges (as evident from AFM image in panel a) or defects. (c) MFM image of post-stress magnetization states (after electrode pairs *AA'* and *BB'* had been sequentially activated). The pinned nanomagnet and the other one in the top row have not rotated under sequential stress while the two nanomagnets in the bottom row have rotated through 180º. The yellow arrows show the magnetization orientations of the two responsive nanomagnets in the bottom row pre- and post-stress.

Next, one of the electrode pairs (*AA'*) is activated by imposing a voltage of 300 V between it and the grounded bottom of the substrate. This results in an average vertical electric field of 0.6 MV/m along the substrate's thickness underneath the electrodes which generates a highly localized out-of-plane tensile strain due to $d_{33}$ coupling and in-plane compressive strain due to $d_{31}$ coupling. Since the separation between the edges of each pair of electrodes (0.75 mm) is comparable to the substrate thickness, the interaction between the local strain fields generates biaxial strain in the piezoelectric substrate in the region where the nanomagnets are placed[28]. The resultant strain is tensile along the line joining *AA'* and compressive along the direction perpendicular to it (see Fig. S1 in the supplementary material). It will be almost entirely transferred to the nanomagnets since the nanomagnets' thickness is much smaller than that of the substrate. A full strain profile was generated with finite element (FE) analysis and conformed to the above description [see Fig. S1 in the supplementary material].

Following the application of the voltage at electrode pair *AA'*, the same voltage (300 V) is applied to the electrode pair *BB'* and subsequently *AA'* is deactivated. Finally, the pair *BB'* is also deactivated. Fig. 2c shows the post-stress MFM image. The magnetizations of the two nanomagnets in the top row have not changed under stress. One of them (at left) is stuck in the pinned state that even a 0.2 Tesla magnetic field could not unpin. The other (at right) could also have been pinned or did not rotate through 180$^o$ owing to insufficient strain transfer. However, the remaining two nanomagnets in the bottom row have flipped their magnetizations (180$^o$ rotation) after the application of stress. In the supplementary material, we discuss why and how some nanomagnets could experience insufficient strain. We can discount the top left nanomagnet whose magnetization is strongly pinned, in which case two out of three nanomagnets underwent complete magnetization reversal.

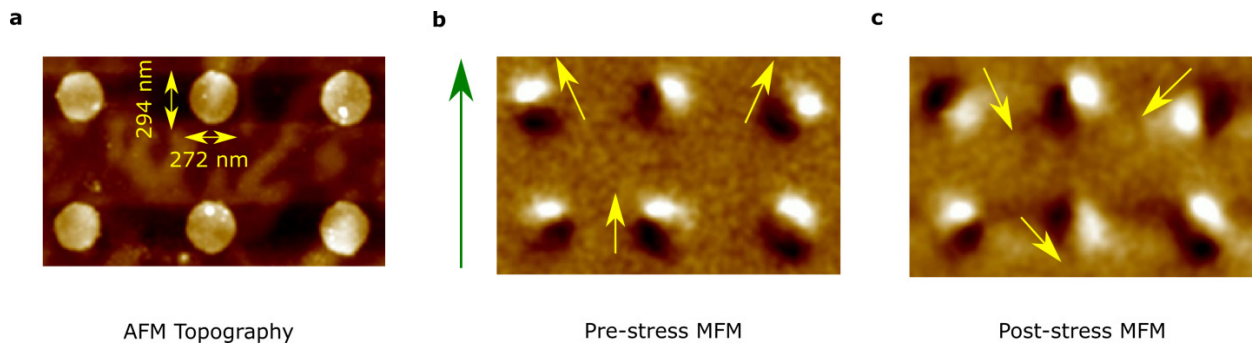

Fig. 3: **AFM and MFM images of six nominally identical larger elliptical Co nanomagnets delineated on a PMN-PT substrate showing how their magnetizations evolve with stress**. (a) AFM image showing the topography of the six nanomagnets. (b) MFM image showing the magnetization states of the nanomagnets after magnetizing them with a ~0.2 Tesla magnetic field along the direction of the green arrow and before subjecting them to stress. (c) MFM image of the nanomagnets after subjecting them to sequential stress along two directions by sequentially activating the electrode pairs. The magnetizations of the two peripheral nanomagnets in the top row have rotated by 180$^o$ from their initial orientations (marked with yellow arrows). The nanomagnet in the middle of the bottom row has rotated by less than 180$^o$ and possibly got trapped into a metastable state. The other three nanomagnets are unresponsive to stress, either because the stress generated in them was insufficient to change their states [see supplementary material for how and why this can happen] or their magnetizations have been pinned by spurious states.

To examine any possible size-dependence of this effect, we experimented with another set of nanomagnets (294 nm major axis, 272 nm minor axis and thickness 8 nm) with the same center-to-center separation of 700 nm between nearest neighbors. These dimensions produce an in-plane shape anisotropy energy barrier of 6.776 eV. Fig. 3a shows the AFM image of 6 nominally identical nanomagnets. MFM images before and after the application of stress are shown in Figs. 3b and 3c, respectively. After initialization with a magnetic field (in the direction of the green arrow along the major axes of the nanomagnets), the magnetizations of none of the six nanomagnets became exactly collinear with the major axes (or magnetic field) as shown in Fig. 3b since, once again, the easy axes did not coincide with the major axes. The post-stress MFM images (see Fig. 3c) reveal that three out of the six nanomagnets (marked by the yellow arrows) have evolved to new magnetic states after experiencing the sequential stress cycle. The angular separation between the initial and final magnetic states is ~180º for two out of these three, but one outlier has rotated by less than 180º, probably owing to trapping into an intervening metastable state that prevented complete magnetization reversal. In any case, there is no substantive difference with the previous set, meaning that both large and small nanomagnets behave in essentially the same way.

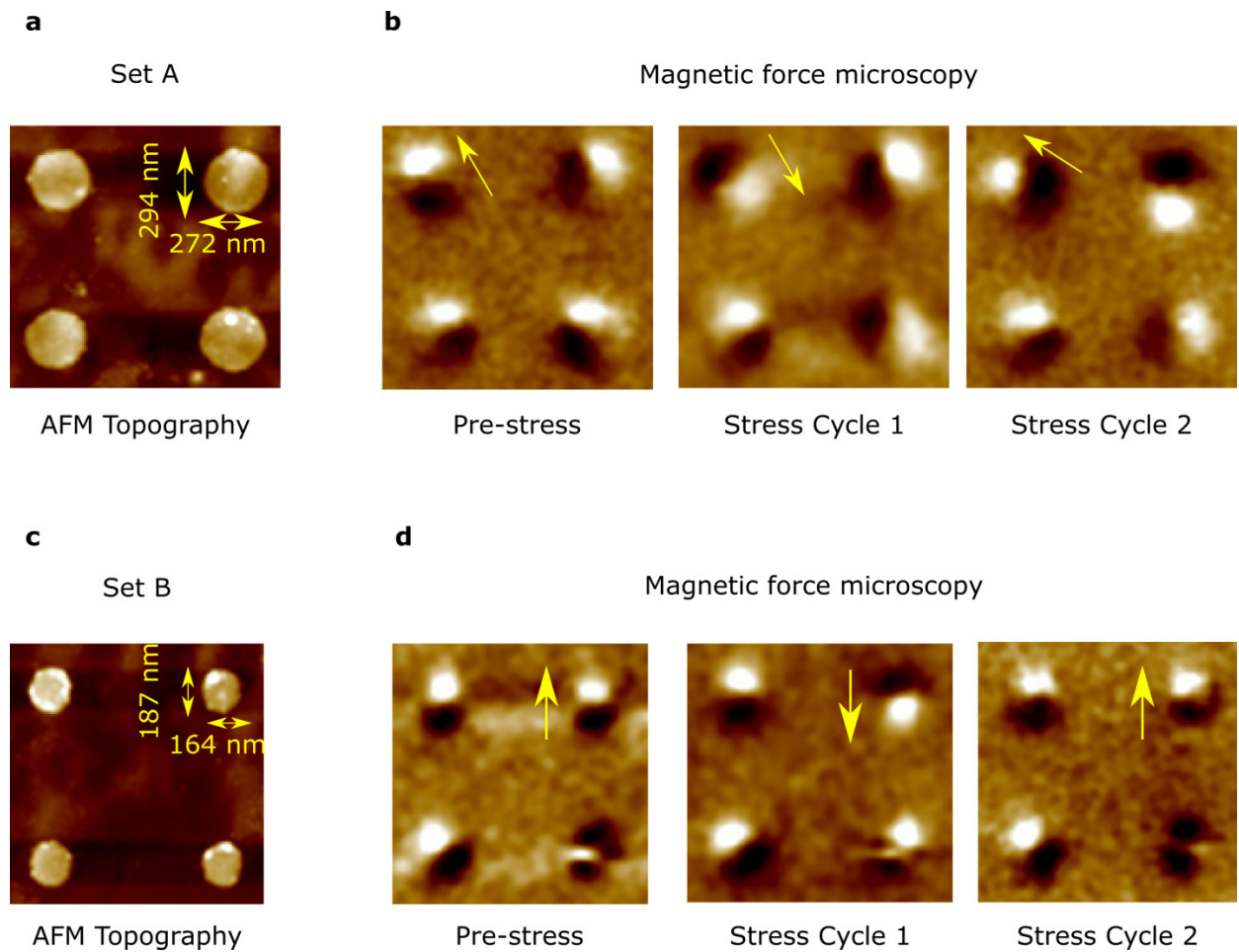

Fig. 4: **AFM and MFM images of two sets of Co nanomagnets (Set A and Set B) delineated on a PMN-PT substrate showing how their magnetizations react to two consecutive cycles of stress**. Set A: 4 nanomagnets with major axis 294 nm and minor axis 272 nm. Set B: 4 nanomagnets with major axis 187 nm and minor axis 164 nm. (**a, c**) AFM image showing the topography of the four isolated nanomagnets. (**b, d**) The left panels show the MFM images of the initial states; the center panels show the MFM image after one sequential stress cycle indicating

that the nanomagnets (marked by yellow arrows in **b** and **d**) experienced complete 180° rotation; the right panels show that the same nanomagnets marked with yellow arrows have undergone another ~180° rotation and hence returned to their initial orientation after the second sequential stress cycle.

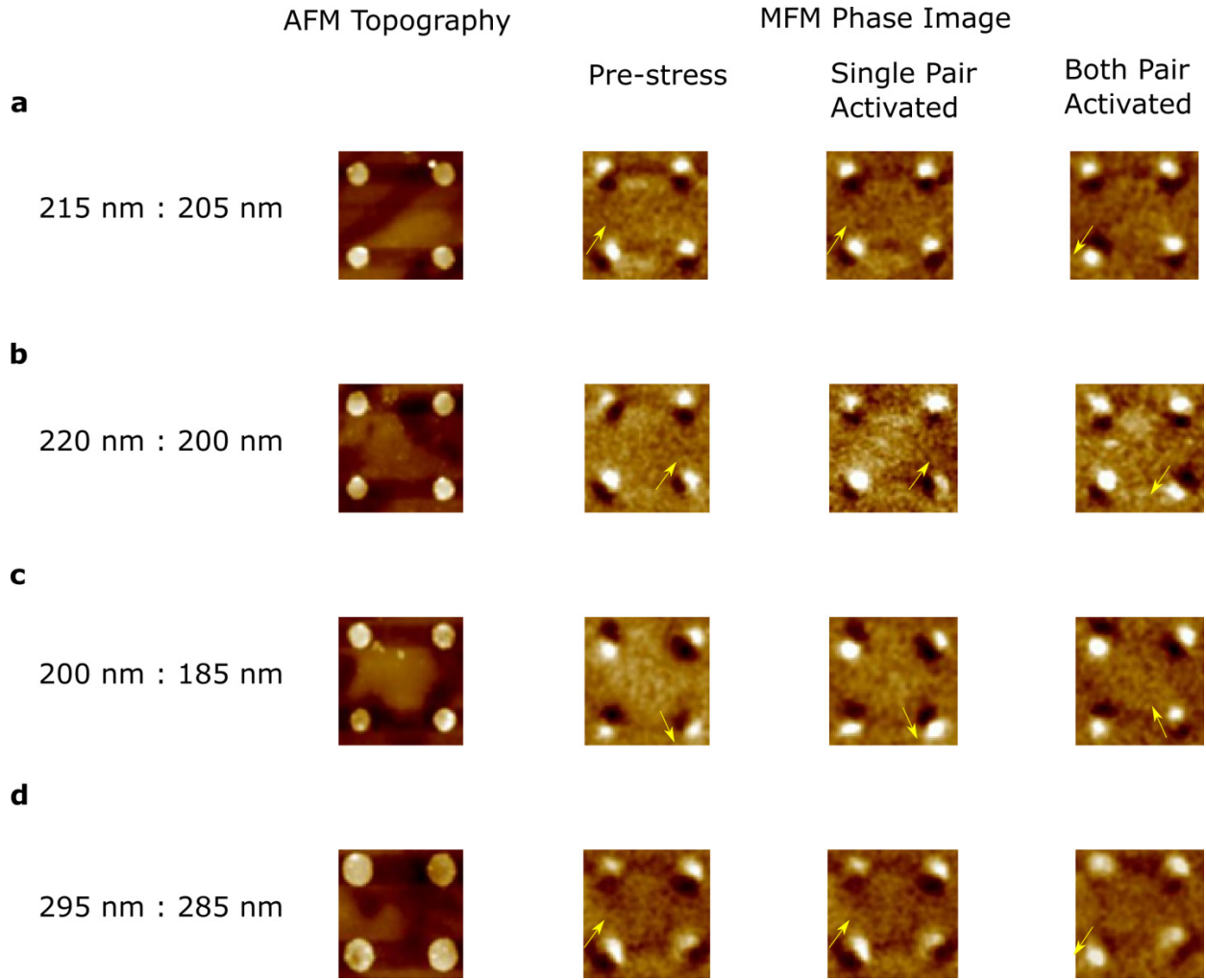

Fig. 5: **AFM and MFM micrographs of four sets of nanomagnets of different sizes and ellipticity showing how their magnetizations evolve when one and both pairs of electrodes are activated**. The nominal dimensions (major and minor axes) are shown on the left in each horizontal panel. The calculated in-plane shape anisotropy energy barriers in these four sets are, respectively, 1.086 eV, 5.728 eV, 4.202 eV and 3.099 eV. The first column shows the topography of the four sets of nanomagnets, the second shows the initial magnetization states after magnetizing with a global magnetic field directed vertically up in this figure, the third shows the magnetization states after one pair of metal pads (say, *AA'*) is activated and then deactivated, while the fourth shows the magnetization states after both pairs are activated successively and deactivated successively.

In order to show that the magnetization of a nanomagnet can be switched *back and forth* between two stable states with consecutive sequential stress cycles, we chose two sets of nanomagnets of thickness 8 nm: Set A with 294 nm major axis and 272 nm minor axis (Fig 4a), and Set B with 187 nm major axis and 164 nm minor axis (Fig 4c). The smaller set has an in-plane shape anisotropy energy barrier of 6.282 eV. Fig. 4b illustrates the switching behavior of Set A nanomagnets in two successive sequential stress cycles. The pre-stress image shows the initial states of four nearest neighbor nanomagnets after magnetization with a magnetic field directed vertically up in this image. After completion of stress cycle 1 (*AA'* and *BB'* sequentially activated *once*), the top left nanomagnet (marked by yellow arrow) has rotated through 180° while the bottom right nanomagnet has evolved into a metastable state whose magnetization has a large angular deviation (almost 90°) from the major axis. The other two nanomagnets are unresponsive to stress (no difference between pre-stress and post-stress). After cycle 2 (*AA'* and *BB'* sequentially activated again), the same top left nanomagnet has rotated by about 160° and returned nearly to the original state, while the unresponsive nanomagnets have remained unresponsive and the one trapped into a metastable state has remained trapped there. Smaller nanomagnets (Set B) behave in the same way, as shown in Fig. 4d, thereby demonstrating size-independence of the magnetization reversal scheme. In this set, the responsive nanomagnet undergoes ~180° rotation in each cycle.

In Fig. 5, we show the stress-induced magnetization changes in four sets of nanomagnets of four different sizes and ellipticity in the four rows. They all have different in-plane shape anisotropy energy barriers, but the generated stress should be able to overcome them in the manner of Fig. 1b and yet the MFM tip should not be able to alter the magnetization. This experiment was carried out to demonstrate that successive activation of *both* electrode pairs is required for magnetization reversal and that activation of only one pair will *not* bring about the 180° rotation. In each row, focus only on the nanomagnet that is marked with a yellow arrow since the other nanomagnets were unresponsive to stress. In the second column, we show the magnetization states of the nanomagnets after magnetization with a 0.2 Tesla global magnetic field in the vertically up direction. In the third column, we show the magnetization states after activating and then deactivating only one pair of electrodes (e. g. *AA'*) and not the other. In the fourth column, we show the magnetization states after sequentially activating and then deactivating both pairs *AA'* and *BB'*. Clearly, activating and deactivating only one pair does not change the state of the nanomagnets in any of the four cases shown (since the second and third columns are always identical). However, activating and deactivating both pairs in succession reverses the magnetization of the responsive nanomagnets in all four cases examined.

What makes this magnetic reversal scheme attractive is the extremely low energy dissipated in the process. The dissipation has three components: the internal dissipation in the nanomagnets which is on the order of 1 aJ at room temperature[32], the mechanical energy dissipation (stress × strain × volume of the nanomagnet) which is also a few aJ, and the electrical energy dissipated in the external circuit to activate the electrodes and generate stress in a nanomagnet. The last component is overwhelmingly dominant and is equal to $CV^2$, where $C$ is the capacitance of the gate pairs and $V$ is the voltage needed to generate the stresses. Let us now estimate how much energy would have been dissipated in a *properly scaled structure* where the separation between a single pair of electrodes is 100 nm, a single nanomagnet of lateral dimension ~50 nm is placed between them, the electrode dimensions are of the same order (100 nm × 100 nm) and the PMN-PT film thickness is 100 nm. Since we applied 300 V across a 0.5 mm piezoelectric substrate to reverse magnetization, we will need to apply only 60 mV across a 100 nm thin film of piezoelectric to see the same effect (i.e., complete 180° rotation). The capacitance of one pair of electrodes will be $C = \varepsilon_r \varepsilon_0 A/d = 2 \times 500 \times 8.854 \times 10^{-12}$ F/m × 100 nm × 100 nm / 100 nm = 0.88 fF, assuming a parallel plate capacitor ($A$ is the area of the electrode, $d$ is the piezoelectric film thickness and $\varepsilon_r$ is the relative dielectric constant of the piezoelectric which is ~500 [40]). The energy dissipated to activate one pair is thus $CV^2 = 2 \times 0.88 \text{ fF} \times (60 \text{ mV})^2 = 6.3$ aJ and to activate both pairs is 13 aJ. This energy would be reduced further if we replace Co with a more magnetostrictive material like FeGa or Terfenol-D. Therefore, the energy dissipated to write one bit of information in a properly scaled memory cell will be a few tens of aJ at worst, which makes this one of the most energy-efficient writing schemes extant. The present experiment establishes a clear pathway to implementing such as scheme and portends an extremely energy-efficient MRAM technology.

One area of concern in these experiments is that the yield was poor (a small fraction of the nanomagnets switched while others were unresponsive to stress), but this can be improved substantially by choosing magnetostrictive materials that have much higher magnetostriction than Co (e.g. Terfenol-D, FeGa) [see supplementary material] and by eliminating pinning sites with careful material preparation.

In conclusion, we have shown complete magnetization reversal in a magnetostrictive elliptical nanomagnet (with bistable magnetization) using mechanical strain generated by sequential electrode activation on a piezoelectric substrate. This can be utilized to write bits in non-volatile memory cells with extremely low energy dissipation. The writing electrodes (four of them) are separate from the reading electrodes (two of them) which has its own advantage but also the disadvantage of a larger cell footprint which reduces the density of memory cells and calls for a more complex architecture. Thus, this scheme is attractive for low-energy, but not high-density, applications.There are many applications in mobile electronics, wearable electronics, space-based computers and medically implanted processors where energy-efficiency outweighs all other considerations and this methodology of writing information in non-volatile memory will be extremely attractive for those applications.

**Methods**

The fabrication process begins by poling a (011)-oriented $Pb(Mg_{1/3}Nb_{2/3})O_3$-$PbTiO_3$ (PMN-PT) [10 mm $\times$10 mm $\times$0.5 mm] (70% PMN and 30\% PT) with an electric field of 0.8 MV/m along the thickness by applying a positive voltage of 400 V on the top polished surface while grounding the bottom surface. Next, two pairs of large area metal pads Ti + Au (10 + 70 nm thickness) with lateral dimensions 0.4 mm $\times$ 0.4 mm are patterned on the surface of the substrate using photolithography and metal evaporation with e-beam, followed by lift-off. The lines joining the centers of opposite pairs subtend an angle of ~60º between them. The distance between the facing edges of the electrodes in each pair is 0.75 mm, comparable to the thickness of the substrate. Next, elliptical Co nanomagnets are fabricated in the space between the electrodes with e-beam lithography (using multilayer PMMA), Co evaporation and lift-off, and aligned such that the major axes of all the nanomagnets subtend angles of approximately +30º and -30º with the two lines joining the centers of the two electrode pairs. To carry out the e-beam lithography, two layers of e-beam resist PMMA – Poly(methyl methacrylate) with different molecular weights (495K PMMA and 950K PMMA; 2% Anisole) were spin coated at 2500 rpm in two steps. The resists were baked at 115º Celsius for 2 minutes in each step. The sample was then exposed under electron beam from a Hitachi SU-70 SEM with a Nabity attachment using an accelerating voltage of 30 kV and a beam current of 60 pA. Subsequently, the resists were developed in MIBK:IPA (1:3) [(methyl isobutyl ketone: isopropyl alcohol)] solution for 90 seconds followed by cold IPA rinse. A Co layer of thickness of ~8 nm is then deposited on the patterned resist using electron beam evaporation at a base pressure of $2 \times 10^{-7}$ Torr with a Ti adhesion layer (thickness of 4 nm). A lift-off process was conducted for removing the resist and metal to produce the final nanomagnet pattern. Finally, the magnets are initialized by magnetizing them at room temperature with a magnetic field of 0.2 Tesla directed along the major axes of the ellipses, using an electromagnet.

# References

1. Biswas, A. K., Bandyopadhyay, S. & Atulasimha, J. Complete magnetization reversal in a magnetostrictive nanomagnet with voltage-generated stress: A reliable energy-efficient non-volatile magneto-elastic memory. *Appl. Phys. Lett*. **105**, 072408 (2014).
2. Slonczewski, J. C. Current-driven excitation of magnetic multilayers. *J. Magn. Magn. Mater.*, **159**, L1-L7 (1996).
3. Berger, L. Emission of spin waves by a magnetic multilayer traversed by a current. *Phys. Rev. B*, **54**, 9353-9358 (1996).
4. Sankey, J. C., *et al*. Measurement of the spin-transfer-torque vector in magnetic tunnel junctions. *Nat. Phys*., **4**, 67-71 (2007).
5. Liu, L., *et al.* Spin-torque switching with the giant spin Hall effect of tantalum. *Science*, **336**, 555-558 (2012).
6. Bhowmik, D., You, L. & Salahuddin, S. Spin Hall effect clocking of nanomagnetic logic without a magnetic field. *Nat. Nanotechnol*., **9**, 59-63 (2013).
7. Mellnik, A. R. *et al*. Spin-transfer torque generated by a topological insulator. *Nature*, **511**, 449-451 (2014).
8. Miron, I. M. *et al*. Fast current-induced domain-wall motion controlled by the Rashba effect. Nat. Mater., **10**, 419-423 (2011).
9. Yu, G. *et al*. Switching of perpendicular magnetization by spin-orbit torques in the absence of external magnetic fields. *Nat. Nanotechnol*., **9**, 1-7 (2014).
10. Fan, Y. *et al*. Magnetization switching through giant spin-orbit torque in a magnetically doped topological insulator heterostructure. *Nat. Mater*., **13**, 699-704 (2014).
11. Fan, Y. *et al*. Electric-field control of spin-orbit torque in a magnetically doped topological insulator. *Nat. Nanotechnol.*, **11**, 352-359 (2016).
12. Yamanouchi, M., Chiba, D., Matsukura, F. & Ohno, H. Current-induced domain wall switching in a ferromagnetic semiconductor structure. *Nature*, **428**, 539-542 (2004).
13. Allwood, D. A. *et al*., Magnetic domain-wall logic. *Science*, **309**, 1688-1692 (2005).
14. Wang, W.-G., Li, M., Hageman, S. & Chien, C. L. Electric-field-assisted switching in magnetic tunnel junctions. *Nat. Mater*., **11**, 64-68 (2011).
15. Chu, Y.-H. *et al*. Electric-field control of local ferromagnetism using a magnetoelectric multiferroic. *Nat. Mater*.. **7**, 478-482 (2008).
16. Heron, J. T. *et al*. Deterministic switching of ferromagnetism at room temperature using an electric field. *Nature*, **516**, 370-373 (2014).
17. Matsukura, F., Tokura, Y. \& Ohno, H. Control of magnetism by electric fields. *Nat. Nanotechnol*.,**10**, 209-220 (2015).
18. Bauer, U. *et al*. Magneto-ionic control of interfacial magnetism. *Nat. Mater*., **14**, 174-181 (2015).
19. Eerenstein, W., Mathur, N. D. & Scott, J. F.  Multiferroic and magnetoelectric materials. *Nature*, **442**, 759-765 (2006).
20. Pertsev, N. A. & Kohlstedt, H. Magnetic tunnel junction on a ferroelectric substrate. *Appl. Phys. Lett*., **95**, 163503 (2009).
21. Atulasimha, J. & Bandyopadhyay, S. Bennett clocking of nanomagnetic logic using multiferroic single-domain nanomagnets. *Appl. Phys. Lett*., **97**, 173105 (2010).
22. Giordano, S., Dusch, Y., Tiercelin, N., Pernod, P. & Preobrazhensky, V. Combined nanomechanical and nanomagnetic analysis of magnetoelectric memories. *Phys. Rev. B*, **85,** 155321 (2012).
23. Biswas, A. K., Bandyopadhyay, S. & Atulasimha, J. Energy-efficient magnetoelastic non-volatile memory. *Appl. Phys. Lett*., **104**, 232403 (2014).
24. Wu, T. *et al*. Electrical control of reversible and permanent magnetization reorientation for magnetoelectric memory devices. *Appl. Phys. Lett*., **98**, 262504 (2011).

25. Li, P. *et al*. Electric field manipulation of magnetization rotation and tunneling magnetoresistance of magnetic tunnel junctions at room temperature. *Adv. Mater*., **26**, 4320-4325 (2014).
26. Fukami, S. *et al*. Low-current perpendicular domain wall motion cell for scalable high-speed MRAM. *2009 IEEE Symp. VLSI Technol. Dig. Tech. Pap*., 230-231 (2009).
27. Amiri, P. K. & Wang, K. L. Voltage-controlled magnetic anisotropy in spintronic devices. *Spin*, **2,** 1240002 (2012).
28. Cui, J. *et al*. A method to control magnetism in individual strain-mediated magnetoelectric islands. *Appl. Phys. Lett.*, **103**, 232905 (2013).
29. Liang, C. Y. *et al*. Electrical control of a single magnetoelastic domain structure on a clamped piezoelectric thin film - Analysis. *J. Appl. Phys.,* **116**, 123909 (2014).
30. Cui, J. *et al*. Generation of localized strain in a thin film piezoelectric to control individual magnetoelectric heterostructures. *Appl. Phys. Lett*., **107**, 092903 (2015).
31. Roy, K., Bandyopadhyay, S. & Atulasimha, J. Hybrid spintronics and straintronics: A magnetic technology for ultralow energy computing and signal processing. *Appl. Phys. Lett*., **99**, 063108 (2011).
32. Roy, K., Bandyopadhyay, S. & Atulasimha, J. Energy dissipation and switching delay in stress-induced switching of multiferroic nanomagnets in the presence of thermal fluctuations. *J. Appl. Phys*., **112**, 023914 (2012).
33. Fashami, M. S., Roy, K., Atulasimha, J. & Bandyopadhyay, S. Magnetization dynamics, Bennett clocking and associated energy dissipation in multiferroic logic. *Nanotechnology*, **22**, 155201 (2011).
34. Ahmad, H., Atulasimha, J. & Bandyopadhyay, S. Reversible strain-induced magnetization switching in FeGa nanomagnets: Pathway to a rewritable, non-volatile, non-toggle, straintronic memory cell for extremely low energy operation. *Sci. Rep*., **5**, 18264 (2015).
35. D'Souza, N., Salehi Fashami, M., Bandyopadhyay, S. & Atulasimha, J. Experimental clocking of nanomagnets with strain for ultralow power Boolean logic. *Nano Lett*., **16,** 1069-1075 (2016).
36. Zhao, Z. *et al*. Giant voltage manipulation of MgO-based magnetic tunnel junctions via localized anisotropic strain: Pathway to ultra-energy-efficient memory technology. *Appl. Phys. Lett*., **109**, 092403 (2016).
37. Roy, K., Bandyopadhyay, S. & Atulasimha, J. Binary switching in a 'symmetric' potential landscape. *Sci. Rep*., **3**, 3038 (2013).
38. Peng, R.-C. *et al*. Fast 180° magnetization switching in a strain-mediated multiferroic heterostructure driven by a voltage. *Sci. Rep*., **6,** 27561 (2016).
39. Wang, J. J. *et al*. Full 180° magnetization reversal with electric fields. *Sci. Rep*., **4**, 7507 (2014).
40. Shaikh, P. A. & Kolekar, Y. D. Study of microstructural, electrical and dielectric properties of perovskite (0.7) PMN-(0.3) PT ferroelectric at different sintering temperatures. *J. Anal. Appl. Pyrolysis*, **93**, 41-46 (2012).

**Acknowledgement**

This work is supported by the US National Science Foundation (NSF) under grant ECCS-1124714 and by the Semiconductor Research Corporation (SRC) under NRI task 2203.001. Additional funding was received from the State of Virginia Commonwealth Research Commercialization Fund under the matching fund grant MF-15-006-MS. J. A's work is also supported by NSF CAREER grant CCF-1253370.

**Author contributions:** A. K. B., J. A. and S. B. conceived and designed the experiments. A. K. B. and H. A. fabricated the structures and made all the measurements. A. K. B., H. A., J. A. and S. B. analyzed the data. All authors contributed to writing the paper.

# SUPPLEMENTARY MATERIAL

**Finite element analysis of the stress profile**

We have carried out finite element analysis of the strain profile generated in the PMN-PT substrate's surface when electrode pairs *AA'* or *BB'* are activated with the applied voltages of 300 V. This is shown in Fig. S1 where the electrodes are the square shaped objects labeled as *A, A', B* and *B'*. The profiles are calculated with COMSOL Multiphysics [1].

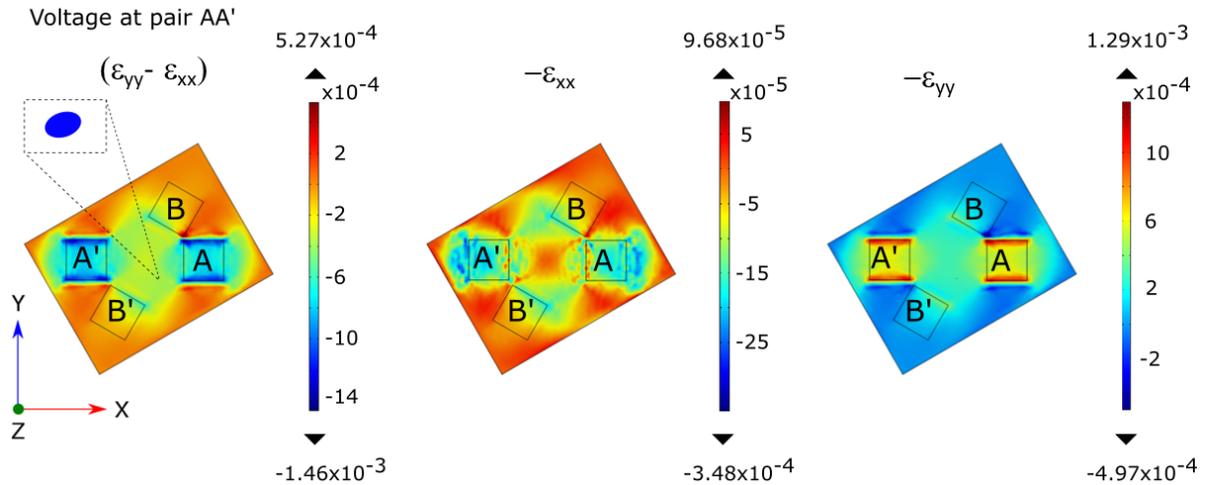

Fig. S1: **Strain profile on the surface of the PMN-PT substrate obtained using finite element analysis with COMSOL Multiphysics package**. The electrode pair *AA'* (horizontal) is activated with a dc voltage of 300 V resulting in a vertical electric field of 0.6 MV/m. An elliptical nanomagnet is drawn in the subfigure at far left to show the alignment of its major and minor axes with respect to the electrode placement. A single ellipse is shown for illustrative purposes, but there are many Co nanomagnets (with their major axis parallel to that of the ellipse shown) in the space between the electrodes.

The strain pattern is clearly complex and approximates a biaxial distribution. Since there is no direct way to incorporate such a complex pattern in micromagnetic simulations, we make an approximation. The net strain ($\varepsilon_{xx} - \varepsilon_{yy}$) is clearly tensile (positive strain) along the line joining the centers of the electrodes in a pair when only one pair is activated. We have verified that it remains tensile when both pairs are activated. Therefore, we will model the effect of activating either pair of electrodes as generation of uniaxial tensile strain of magnitude ($\varepsilon_{xx} - \varepsilon_{yy}$) along the line joining the centers of that pair. When both pairs are activated, we will assume the presence of two uniaxial strains, each of magnitude ($\varepsilon_{xx} - \varepsilon_{yy}$), along the two directions. We make these assumptions since micromagnetic simulators can handle uniaxial strain directly, but cannot handle more complex strain distributions. We will also assume that the strain generated in the substrate is completely transferred to the nanomagnet. These assumptions may affect the quantitative results, but not the broad qualitative features.

**Micromagnetic simulations**

Micromagnetic simulations of magnetization dynamics and magnetic states evolving under stress are carried out with the OOMMF package where the strain is incorporated as described earlier. We assume a constant (spatially invariant) value of the uniaxial strain in the nanomagnet (resulting in a fixed stress anisotropy energy density) that lies between $3\times10^{-4}$ and $4\times10^{-4}$. The strain is assumed to be spatially invariant in order to reduce the (otherwise enormous) computational burden. In the simulations, the saturation magnetization of Co is assumed to be $10^6$ A/m [2] and the exchange stiffness is 21 pJ/m [3]. The Gilbert damping coefficient is assumed to be 0.027 instead of the bulk value of 0.008 to account for surface states in nanomagnets which increase the damping rate. The initial state is computed assuming that an applied magnetic field of 0.2 Tesla was directed up along the major axis of the nanomagnet, which was removed prior to initiating the stress cycle. The electrode voltages are assumed to be pulsed with zero rise time and the stress is assumed to respond instantaneously to the voltage. These idealizations do not affect the qualitative features of the pre- and post-stress magnetization states. The voltage timing diagram is shown at the top of the center panel in Fig. S2, which also shows the results of micromagnetic simulation for three different nanomagnet sizes: large (294 nm major axis and 272 nm minor axis), intermediate (198 nm major axis and 183 nm minor axis) and small (187 nm major axis and 164 nm minor axis). Although the nominal thickness of the deposited Co layer is ~8 nm, we assumed the effective thickness to be 7 nm to allow for oxidation of the Co surface down to 1 nm. Fig. S2 confirms successful magnetization reversal under sequential stress application in all cases.

**Stress calculation**

In the main text, we stated that the strain generated in the surface of the PMN-PT substrate (underneath the nanomagnet's center) is between $3\times10^{-4}$ and $4\times10^{-4}$. Let us verify if this is enough to switch the nanomagnets. The stress anisotropy energy is given by

$$E_{str} = K_{us}\Omega \qquad (S1)$$

where $K_{us} = -\dfrac{3}{2}\lambda_s\sigma$ is the stress anisotropy energy density, $\Omega$ is the volume of the nanomagnet, $\lambda_s$ is the magnetostriction coefficient and $\sigma$ is the stress. We will assume that the strain is small enough that Hooke's law is valid and hence $\sigma = \varepsilon Y$ where $\varepsilon$ is the strain and $Y$ is the Young's modulus of Co.

The value of $\lambda_s$ in a Co thin film can vary from sample to sample between -20 and -60 ppm [4]. The strain profile in Fig. S1 shows that the strain varies between $3\times10^{-4}$ and $4\times10^{-4}$. Based on these data, we can estimate the stress anisotropy energy density $K_{us}$ generated in the nanomagnets when the electrode pairs are activated and the strain varies between $3\times10^{-4}$ and $4\times10^{-4}$. The Young's modulus of Co is 209 GPa. Therefore, the lower and upper limit of $K_{us}$ will be

$$K_{us-lower} = \frac{3}{2}\lambda_{s-low}\varepsilon_{low}Y = \frac{3}{2} \times 20 \times 10^{-6} \times 3 \times 10^{-4} \times 209 \times 10^{9} = 1881 \text{ J/m}^3$$

$$K_{us-upper} = \frac{3}{2}\lambda_{s-high}\varepsilon_{high}Y = \frac{3}{2} \times 60 \times 10^{-6} \times 4 \times 10^{-4} \times 209 \times 10^{9} = 7524 \text{ J/m}^3$$

(S2)

**a**

294 nm : 272 nm : 7 nm

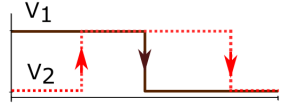

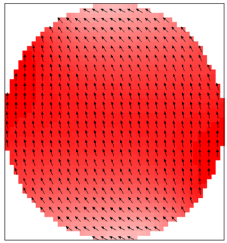

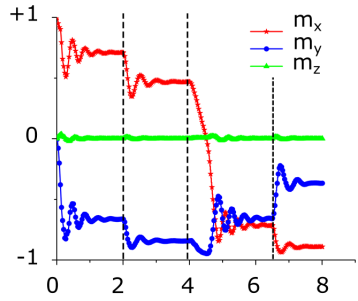

$K_{us}$ = 3200 J/m³

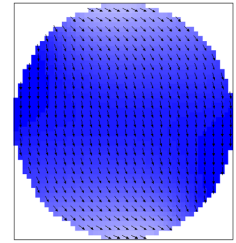

**b**

198 nm : 183 nm : 7 nm

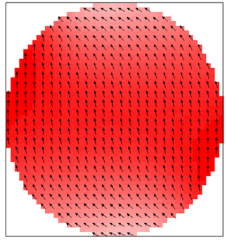

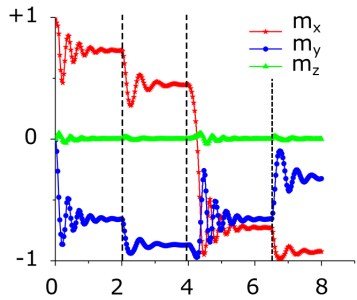

$K_{us}$ = 4800 J/m³

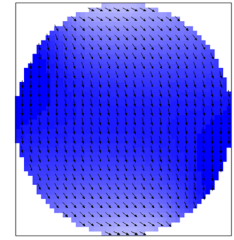

**c**

187 nm : 164 nm : 6 nm

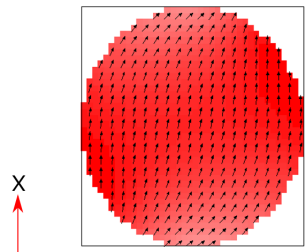

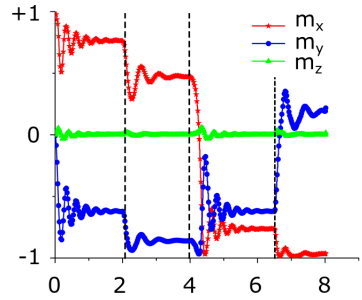

$K_{us}$ = 6400 J/m³

Pre-stress        Time (ns)        Post-stress

Fig. S2: **Magnetization dynamics and the initial and final spin textures of three nanomagnets with dimensions (294 nm: 272 nm: 7 nm), (198 nm: 183 nm: 7 nm) and (187 nm: 164 nm: 6 nm) subjected to stress**. These results are obtained using OOMMF. The stress timing diagram is shown at the top of the middle panel. In order to keep the simulation tractable, the stress pulse widths are assumed to be 2 ns in the simulation even though in the actual experiments, the voltages at the electrodes were turned on and off manually in time scales of several seconds. Since steady state is nearly achieved in 2 ns, there is no substantive difference between the results for 2 ns and several seconds. (Left) The initial magnetization state showing the spin orientations in the nanomagnet (red represents one direction along the major axis and blue the anti-parallel direction). (Right) The final magnetization state. (Middle) All three components of magnetization are plotted as a function of time. Note that stress anisotropy energy density $K_{us}$ required to switch the nanomagnets varies from 3200 J/m$^3$ for the largest nanomagnets to 6400 J/m$^3$ for the smallest nanomagnets.

Our micromagnetic simulations show that the minimum value of $K_{us}$ required to switch the largest nanomagnets used in the experiments is 3200 J/m$^3$ and for the smallest nanomagnets is 6400 J/m$^3$. The calculated in-plane shape anisotropy energy *barrier per unit volume* for the three different sets are (from largest to smallest): 2481 J/m$^3$, 3387 J/m$^3$ and 5998 J/m$^3$. The stress anisotropy energy density $K_{us}$ required to switch the magnetization should increase with the value of the in-plane shape anisotropy energy barrier per unit volume because switching requires stress to overcome the in-plane shape anisotropy barrier in the manner of Fig. 1b. That is why the minimum $K_{us}$ required for switching is highest for the smallest nanomagnet and lowest for the largest.

The range of $K_{us}$ required to switch the nanomagnets used in the experiments (3200 -- 6400 J/m$^3$) is *within* the calculated lower and upper bounds in Equation (S2), which means that sufficient strain could have been generated in the experiments to make the nanomagnets switch. Hence, some nanomagnets will switch, *but not all*. The largest nanomagnets which experience $K_{us}$ between 1881 J/m$^3$ and 3200 J/m$^3$ will not switch because of insufficient strain. Similarly, the smallest nanomagnets which experience $K_{us}$ between 1881 J/m$^3$ and 6400 J/m$^3$ will also not switch because of insufficient strain. That is why some nanomagnets fail to switch. Lithographic imperfections, resulting in a variation of the nanomagnet size, will further increase the failure probability. The jagged edges of the nanomagnets may spawn metastable states where the magnetizations can end up getting pinned, which will increase the failure probability even further. All these effects conspire to make the switching probability small, which is why we observe a small fraction of the nanomagnets undergoing complete magnetic reversal. The best way to increase the switching probability is to use materials like Terfenol-D that have ~30 times higher magnetostriction than Co and therefore experience much higher $K_{us}$ for a given stress value. This remains a goal for future experiments.

**Magnetization states at various stages of the sequential stress cycle computed with OOMMF**

In Fig. S3, we show the magnetization states of different sized nanomagnets at different stages of the stressing cycle computed with OOMMF. When both pairs are activated, we assume two uniaxial stresses directed along the axes joining the centers of the pairs and when only one pair is activated, we assume a single uniaxial stress directed along the axis joining the centers of that pair.

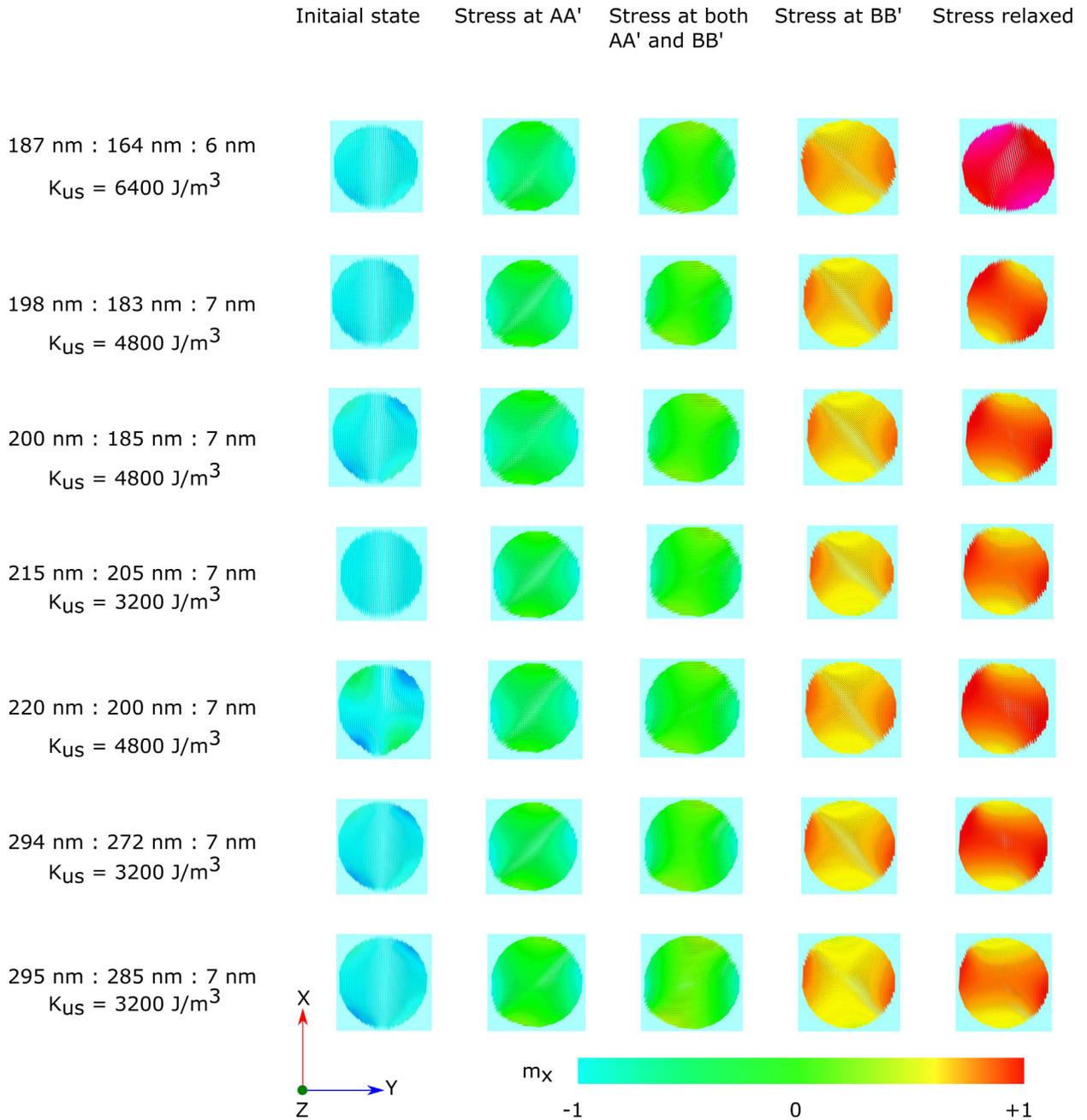

Fig. S3: **Magnetization states of nanomagnets of different sizes at various stages of the sequential stressing cycle calculated with OOMMF**. Cyan represents magnetization pointing up along the nanomagnets' major axes and red represents magnetization pointing down, with other colors representing intermediate orientations.

**The role of the Gilbert damping in magnetization reversal**

In our micromagnetic simulations, we assumed the value of the Gilbert damping factor $\alpha$ to be 0.027 which is higher than the value reported in bulk cobalt (0.008) [2]. The rationale for assuming a higher value of $\alpha$ is that $\alpha$ should be higher in nanomagnets than in bulk because of the influence of surface states. Nanomagnets have much higher surface to volume ratio than bulk and hence will be more susceptible to surface states. In strain mediated switching, α determines how fast the magnetization would settle down to the nearest energy minimum, but it does not affect how much stress is needed to switch. This is in contrast to the spin-transfer-torque (STT) assisted switching mechanism where the critical current for switching is proportional to α. Nonetheless, in order to confirm that the qualitative features of the magnetic reversal are insensitive to α, we carried out a simulation with α=0.008. The results are shown in Fig. S4. There are more oscillations, or "ringing", in the magnetization dynamics because of the lower damping, and it takes longer for the nanomagnets to transition between states, but the qualitative features of the magnetization reversal (180° rotation) are unchanged.

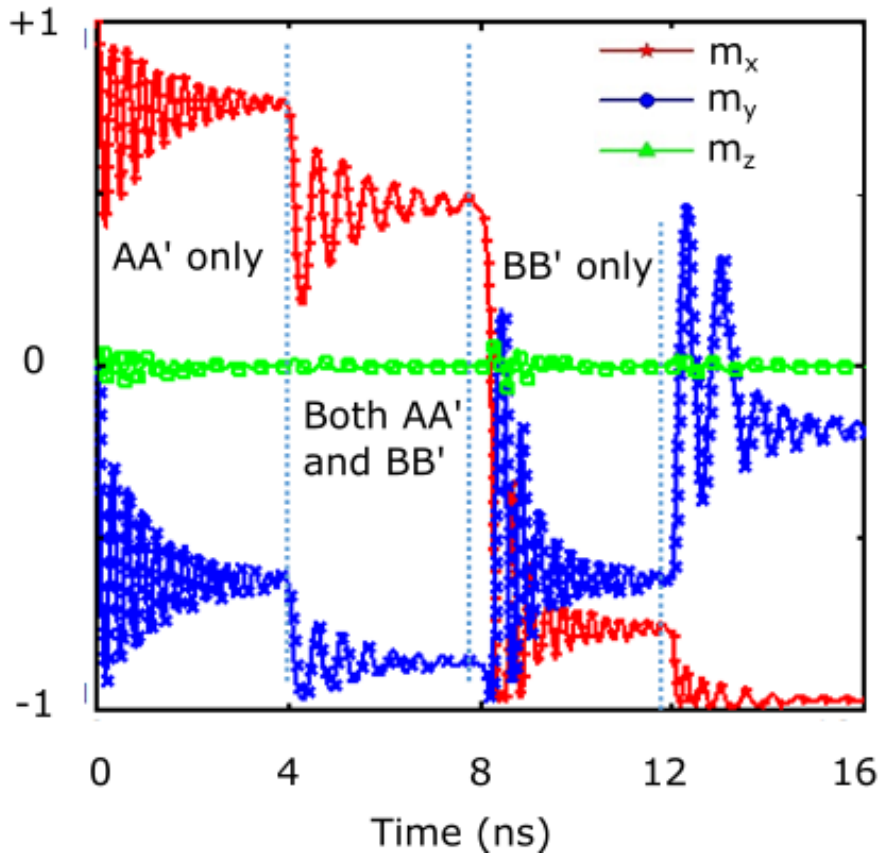

Fig. S4: **Magnetization dynamics of a Co nanomagnet with dimensions (187 nm: 164 nm: 6 nm) under the sequential stressing cycle**. The value of the Gilbert damping coefficient here is 0.008 instead of 0.027. The lower damping causes more ringing and longer transition times, but does not inhibit the complete magnetization reversal or 180° rotation.

**References**

1. M. Shanthi, L. C. Lim, K. K. Rajan & J. Jin. Complete sets of elastic, dielectric, and piezoelectric properties of flux-grown [011]- poled Pb(Mg$_{1/3}$Nb$_{2/3}$)O$_3$-PbTiO$_3$ single crystals. *Appl. Phys. Lett*. **92**, 142906 (2008).
2. M. Tokaç, S. A. Bunyaev, G. N. Kakazei, D. S. Schmool, D. Atkinson & A. T. Hindmarch. Interfacial Structure Dependent Spin Mixing Conductance in Cobalt Thin Films. *Phys. Rev. Lett*., **115**, 1–5 (2015).
3. C. Eyrich, *et al*., Exchange Stiffness in Thin-Film Cobalt Alloys. *J. Appl. Phys*., **111**, 07C919 (2012).
4. E. Klokholm and J. Aboaf. The saturation magnetostriction of thin polycrystalline films of iron, cobalt, and nickel. *J. Appl. Phys*., **53**, 2661–2663 (1982).